\documentclass[fleqn,vecphys,sectrefs]{svmono}
\usepackage{graphicx}
\usepackage{amsmath}
\usepackage{cite}

\input{tcilatex}

\begin{document}

\author{K. Stali\={u}nas and V.J. S\'{a}nchez-Morcillo}
\title{Transverse Patterns in \\
Nonlinear Optical Resonators\\
{\small \ Springer Tracts in Modern Physics, vol. 183.}\\
{\small 2003, 226 p. with 132 figures. Hardcover. }\\
{\small ISBN 3-540-00434-3}}
\maketitle

\begin{center}
\vspace{0.75cm}

{\LARGE Abstract}\vspace{0.75cm}
\end{center}

This book is devoted to the formation and dynamics of localized structures
(vortices, solitons) and of extended patterns (stripes, hexagons, tilted
waves) in nonlinear optical resonators such as lasers, optical parametric
oscillators, and photorefractive oscillators. Theoretical analysis is
performed by deriving order parameter equations, and also through numerical
integration of microscopic models of the systems under investigation.
Experimental observations, and possible technological implementations of
transverse optical patterns are also discussed. A comparison with patterns
found in other nonlinear systems, i.e. chemical, biological, and
hydrodynamical systems, is given.\vspace{1cm}

 \vspace{1cm}

The first chapter of the book (the introduction) and the table of contents
are given in this article. The full text of the book is available at:\vspace{%
1cm}

http://springeronline.com \vspace{0.75cm}

\begin{figure}[h]
\includegraphics[width=.3\textwidth]{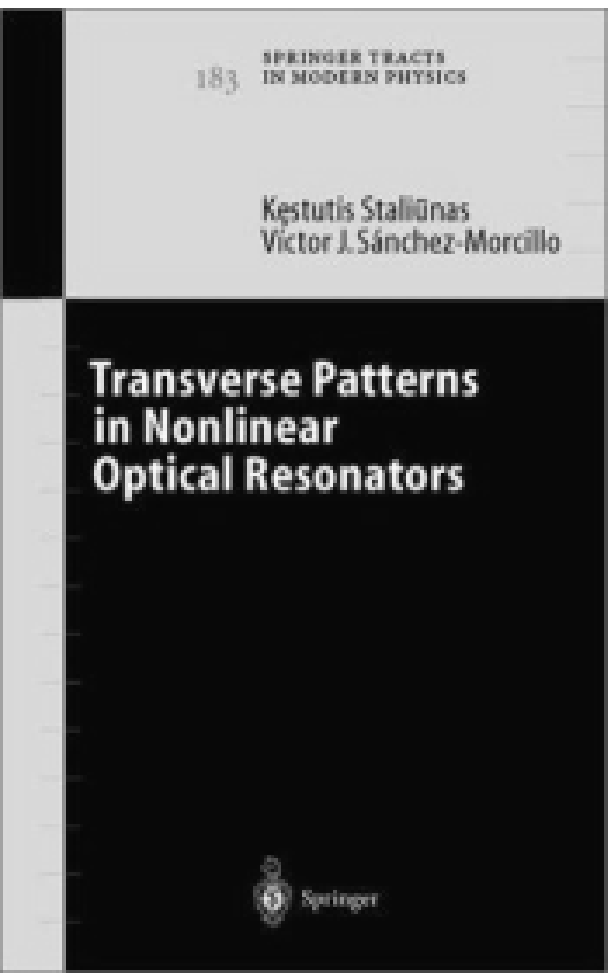}
\end{figure}
\pagebreak

\section*{Table of Contents}

\contentsline {chapter}{\numberline {1}Introduction}{ } 
\contentsline
{section}{\numberline {1.1}Historical Survey}{ } 
\contentsline
{section}{\numberline {1.2}Patterns in Nonlinear Optical Resonators}{ } 
\contentsline {subsection}{\numberline {1.2.1}Localized Structures: Vortices
and Solitons}{ } 
\contentsline {subsection}{\numberline {1.2.2}Extended
Patterns}{ } 
\contentsline {section}{\numberline {1.3}Optical Patterns in
Other Configurations}{ } 
\contentsline {subsection}{\numberline
{1.3.1}Mirrorless Configuration}{ } 
\contentsline {subsection}{\numberline
{1.3.2}Single-Feedback-Mirror Configuration}{ } 
\contentsline
{subsection}{\numberline {1.3.3}Optical Feedback Loops}{ } 
\contentsline
{section}{\numberline {1.4}The Contents of this Book}{ } 
\contentsline
{section}{References}{ } 
\contentsline {chapter}{\numberline {2}Order
Parameter Equations for Lasers}{ } 
\contentsline {section}{\numberline
{2.1}Model of a Laser}{ } 
\contentsline {section}{\numberline {2.2}Linear
Stability Analysis}{ } 
\contentsline {section}{\numberline {2.3}Derivation
of the Laser Order Parameter Equation}{ } 
\contentsline
{subsection}{\numberline {2.3.1}Adiabatic Elimination}{ } 
\contentsline
{subsection}{\numberline {2.3.2}Multiple-Scale Expansion}{ } 
\contentsline
{section}{References}{ } 
\contentsline {chapter}{\numberline {3}Order
Parameter Equations \linebreak for Other Nonlinear Resonators}{ } %
\contentsline {section}{\numberline {3.1}Optical Parametric Oscillators}{ } 
\contentsline {section}{\numberline {3.2}The Real Swift--Hohenberg Equation
for DOPOs}{ } 
\contentsline {subsection}{\numberline {3.2.1}Linear
Stability Analysis}{ } 
\contentsline {subsection}{\numberline
{3.2.2}Scales}{ } 
\contentsline {subsection}{\numberline {3.2.3}Derivation
of the OPE}{ } \contentsline {subsubsection}{(a) Moderate Pump Detuning}{ } %
\contentsline {subsubsection}{(b) Large Pump Detuning}{ } 
\contentsline
{section}{\numberline {3.3}The Complex Swift--Hohenberg Equation for
OPOs}{ } 
\contentsline {subsection}{\numberline {3.3.1}Linear Stability
Analysis}{ } \contentsline {subsection}{\numberline {3.3.2}Scales}{ } %
\contentsline {subsection}{\numberline {3.3.3}Derivation of the OPE}{ } 
\contentsline {section}{\numberline {3.4}The Order Parameter
Equation\linebreak \ for Photorefractive Oscillators}{ } 
\contentsline
{subsection}{\numberline {3.4.1}Description and Model}{ } 
\contentsline
{subsection}{\numberline {3.4.2}Adiabatic Elimination and Operator
Inversion}{ } 
\contentsline {section}{\numberline {3.5}Phenomenological
Derivation\linebreak \ of Order Parameter Equations}{ } 
\contentsline
{section}{References}{ } 
\contentsline {chapter}{\numberline {4}Zero
Detuning: Laser Hydrodynamics\linebreak \ and Optical Vortices}{ } %
\contentsline {section}{\numberline {4.1}Hydrodynamic Form}{ } 
\contentsline
{section}{\numberline {4.2}Optical Vortices}{ } 
\contentsline
{subsection}{\numberline {4.2.1}Strong Diffraction}{ } 
\contentsline
{subsection}{\numberline {4.2.2}Strong Diffusion}{ } 
\contentsline
{subsection}{\numberline {4.2.3}Intermediate Cases}{ } 
\contentsline
{section}{\numberline {4.3}Vortex Interactions}{ } 
\contentsline
{section}{References}{ } 
\contentsline {chapter}{\numberline {5}Finite
Detuning: Vortex Sheets\linebreak \ and Vortex Lattices}{ } 
\contentsline
{section}{\numberline {5.1}Vortices ``Riding'' on Tilted Waves}{ } %
\contentsline {section}{\numberline {5.2}Domains of Tilted Waves}{ } %
\contentsline {section}{\numberline {5.3}Square Vortex Lattices}{ } %
\contentsline {section}{References}{ } 
\contentsline {chapter}{\numberline
{6}Resonators with Curved Mirrors}{ } 
\contentsline {section}{\numberline
{6.1}Weakly Curved Mirrors}{ } 
\contentsline {section}{\numberline
{6.2}Mode Expansion}{ } 
\contentsline {subsection}{\numberline
{6.2.1}Circling Vortices}{ } 
\contentsline {subsection}{\numberline
{6.2.2}Locking of Transverse Modes}{ } 
\contentsline {section}{\numberline
{6.3}Degenerate Resonators}{ } \contentsline {section}{References}{ } %
\contentsline {chapter}{\numberline {7}The Restless Vortex}{ } 
\contentsline
{section}{\numberline {7.1}The Model}{ } 
\contentsline
{section}{\numberline {7.2}Single Vortex}{ } 
\contentsline
{section}{\numberline {7.3}Vortex Lattices}{ } 
\contentsline
{subsection}{\numberline {7.3.1}``Optical'' Oscillation Mode}{ } 
\contentsline {subsection}{\numberline {7.3.2}Parallel translation of a
vortex lattice}{ } 
\contentsline {section}{\numberline {7.4}Experimental
Demonstration of the ``Restless'' Vortex}{ } 
\contentsline
{subsection}{\numberline {7.4.1}Mode Expansion}{ } 
\contentsline
{subsection}{\numberline {7.4.2}Phase-Insensitive Modes}{ } 
\contentsline
{subsection}{\numberline {7.4.3}Phase-Sensitive Modes}{ } 
\contentsline
{section}{References}{ } 
\contentsline {chapter}{\numberline {8}Domains
and Spatial Solitons}{ } 
\contentsline {section}{\numberline
{8.1}Subcritical Versus Supercritical Systems}{ } 
\contentsline
{section}{\numberline {8.2}Mechanisms Allowing Soliton Formation}{ } 
\contentsline {subsection}{\numberline {8.2.1}Supercritical Hopf
Bifurcation}{ } 
\contentsline {subsection}{\numberline {8.2.2}Subcritical
Hopf Bifurcation}{ } 
\contentsline {subsubsection}{(a) The Quintic Complex
Swift--Hohenberg Equation.}{ } 
\contentsline {subsubsection}{(b) Spatial
Nonlinear Resonance.}{ } 
\contentsline {section}{\numberline
{8.3}Amplitude and Phase Domains}{ } 
\contentsline {section}{\numberline
{8.4}Amplitude and Phase Spatial Solitons}{ } 
\contentsline
{section}{References}{ } 
\contentsline {chapter}{\numberline
{9}Subcritical Solitons I: Saturable Absorber}{ } 
\contentsline
{section}{\numberline {9.1}Model and Order Parameter Equation}{ } 
\contentsline {section}{\numberline {9.2}Amplitude Domains and Spatial
Solitons}{ } 
\contentsline {section}{\numberline {9.3}Numerical
Simulations}{ } 
\contentsline {subsection}{\numberline {9.3.1}Soliton
Formation}{ } 
\contentsline {subsection}{\numberline {9.3.2}Soliton
Manipulation: Positioning, Propagation, Trapping and Switching}{ } %
\contentsline {section}{\numberline {9.4}Experiments}{ } 
\contentsline
{section}{References}{ } 
\contentsline {chapter}{\numberline
{10}Subcritical Solitons II: \linebreak Nonlinear Resonance}{ } 
\contentsline {section}{\numberline {10.1}Analysis of the Homogeneous State.
\linebreak Nonlinear Resonance}{ } 
\contentsline {section}{\numberline
{10.2}Spatial Solitons}{ } 
\contentsline {subsection}{\numberline
{10.2.1}One-Dimensional Case}{ } 
\contentsline {subsection}{\numberline
{10.2.2}Two-Dimensional Case}{ } \contentsline {section}{References}{ } 
\contentsline {chapter}{\numberline {11}Phase Domains and Phase
Solitons}{ } 
\contentsline {section}{\numberline {11.1}Patterns in Systems
with a Real-Valued Order Parameter}{ } 
\contentsline {section}{\numberline
{11.2}Phase Domains}{ } 
\contentsline {section}{\numberline {11.3}Dynamics
of Domain Boundaries}{ } 
\contentsline {subsection}{\numberline
{11.3.1}Variational Approach}{ } 
\contentsline {subsection}{\numberline
{11.3.2}Two-Dimensional Domains}{ } 
\contentsline {section}{\numberline
{11.4}Phase Solitons}{ } 
\contentsline {section}{\numberline
{11.5}Nonmonotonically Decaying Fronts}{ } 
\contentsline
{section}{\numberline {11.6}Experimental Realization of Phase
Domains\linebreak \ and Solitons}{ } 
\contentsline {section}{\numberline
{11.7}Domain Boundaries and Image Processing}{ } 
\contentsline
{section}{References}{ } 
\contentsline {chapter}{\numberline {12}Turing
Patterns in Nonlinear Optics}{ } 
\contentsline {section}{\numberline
{12.1}The Turing Mechanism in Nonlinear Optics}{ } 
\contentsline
{section}{\numberline {12.2}Laser with Diffusing Gain}{ } 
\contentsline
{subsection}{\numberline {12.2.1}General Case}{ } 
\contentsline
{subsection}{\numberline {12.2.2}Laser with Saturable Absorber}{ } 
\contentsline {subsection}{\numberline {12.2.3}Stabilization of Spatial
Solitons by Gain Diffusion}{ } 
\contentsline {section}{\numberline
{12.3}Optical Parametric Oscillator \linebreak with Diffracting Pump}{ } 
\contentsline {subsection}{\numberline {12.3.1}Turing Instability in a
DOPO}{ } 
\contentsline {subsection}{\numberline {12.3.2}Stochastic
Patterns}{ } 
\contentsline {subsection}{\numberline {12.3.3}Spatial
Solitons Influenced by Pump Diffraction}{ } 
\contentsline
{section}{References}{ } 
\contentsline {chapter}{\numberline
{13}Three-Dimensional Patterns}{ } 
\contentsline {section}{\numberline
{13.1}The Synchronously Pumped DOPO}{ } 
\contentsline
{subsection}{\numberline {13.1.1}Order Parameter Equation}{ } 
\contentsline {section}{\numberline {13.2}Patterns Obtained from the 3D
Swift--Hohenberg Equation}{ } 
\contentsline {section}{\numberline
{13.3}The Nondegenerate OPO}{ } 
\contentsline {section}{\numberline
{13.4}Conclusions}{ } 
\contentsline {subsection}{\numberline
{13.4.1}Tunability of a System with a Broad Gain Band}{ } 
\contentsline
{subsection}{\numberline {13.4.2}Analogy Between 2D and 3D Cases}{ } %
\contentsline {section}{References}{ } 
\contentsline {chapter}{\numberline
{14}Patterns and Noise}{ } 
\contentsline {section}{\numberline {14.1}Noise
in Condensates}{ } 
\contentsline {subsection}{\numberline
{14.1.1}Spatio-Temporal Noise Spectra}{ } 
\contentsline
{subsubsection}{Spatial Power Spectra.}{ } 
\contentsline
{subsubsection}{Temporal Power Spectra.}{ } 
\contentsline
{subsection}{\numberline {14.1.2}Numerical Results}{ } 
\contentsline
{subsubsection}{Temporal Power Spectra.}{ } 
\contentsline
{subsubsection}{Spatial Power Spectra.}{ } 
\contentsline
{subsection}{\numberline {14.1.3}Consequences}{ } 
\contentsline
{section}{\numberline {14.2}Noisy Stripes}{ } 
\contentsline
{subsection}{\numberline {14.2.1}Spatio-Temporal Noise Spectra}{ } %
\contentsline {subsection}{\numberline {14.2.2}Stochastic Drifts}{ } %
\contentsline {subsection}{\numberline {14.2.3}Consequences}{ } %
\contentsline {section}{References}{ } \contentsline {chapter}{Index}{ }

\chapter{Introduction}

Pattern formation, i.e. the spontaneous emergence of spatial order, is a
widespread phenomenon in nature, and also in laboratory experiments.
Examples can be given from almost every field of science, some of them very
familiar, such as fingerprints, the stripes on the skin of a tiger or zebra,
the spots on the skin of a leopard, the dunes in a desert, and some others
less evident, such as the convection cells in a fluid layer heated from
below, and the ripples formed in a vertically oscillated plate covered with
sand \cite{Cross93}.

All these patterns have something in common: they arise in spatially
extended, dissipative systems which are driven far from equilibrium by some
external stress. ``Spatially extended'' means that the size of the system
is, at least in one direction, much larger than the characteristic scale of
the pattern, determined by its wavelength. The dissipative nature of the
system implies that spatial inhomogeneities disappear when the external
stress is weak, and the uniform state of the system is stable. As the stress
is increased, the uniform state becomes unstable with respect to spatial
perturbations of a given wavelength. In this way, the system overcomes
dissipation and the state of the system changes abruptly and qualitatively
at a critical value of the stress parameter. The very onset of the
instability is, however, a linear process. The role of nonlinearity is to
select a concrete pattern from a large number of possible patterns.

These ingredients of pattern-forming systems can be also found in many
optical systems (the most paradigmatic example is the laser), and,
consequently, formation of patterns of light can also be expected. In
optics, the mechanism responsible for pattern formation is the interplay
between diffraction, off-resonance excitation and nonlinearity. Diffraction
is responsible for spatial coupling, which is necessary for the existence of
nonhomogeneous distributions of light.

Some patterns found in systems of very different nature (hydrodynamic,
chemical, biological or other) look very similar, while other patterns show
features that are specific to particular systems. The following question
then naturally arises: which peculiarities of the patterns are typical of
optics only, and which peculiarities are generic? At the root of any
universal behavior of pattern-forming systems lies a common theoretical
description, which is independent of the system considered. This common
behavior becomes evident after the particular microscopic models have been
reduced to simpler models, called order parameter equations (OPEs). There is
a very limited number of universal equations which describe the behavior of
a system in the vicinity of an instability; these allow understanding of the
patterns in different systems from a unified point of view.

The subject of this book is transverse light patterns in nonlinear optical
resonators, such as broad-aperture lasers, photorefractive oscillators and
optical parametric oscillators. This topic has already been reviewed in a
number of works \cite
{Lugiato92,Weiss92,Lugiato94a,Lugiato98,Arecchi99,Vasnetsov,Kivshar00,Neubecker99,Vorontsov}%
. We treat the problem here by means of a description of the optical
resonators by order parameter equations, reflecting the universal properties
of optical pattern formation.

\section{Historical Survey}

The topic of optical pattern formation became a subject of interest in the
late 1980s and early 1990s. However, some hints of spontaneous pattern
formation in broad-aperture lasers can be dated to two decades before, when
the first relations between laser physics and fluids/superfluids were
recognized \cite{Graham70}. The laser--fluid connection was estalished by
reducing the laser equations for the class A case (i.e. a laser in which the
material variables relax fast compared with the field in the optical
resonator) to the complex Ginzburg--Landau (CGL) equation, used to describe
superconductors and superfluids. In view of this common theoretical
description, it could then be expected that the dynamics of light in lasers
and the dynamics of superconductors and superfluids would show identical
features.

In spite of this insight, the study of optical patterns in nonlinear
resonators was abandoned for a decade, and the interest of the optical
community turned to spatial effects in the unidirectional mirrorless
propagation of intense light beams in nonlinear materials. In the simplest
cases, the spatial evolution of the fields is just a filamentation of the
light in a focusing medium; in more complex cases, this evolution leads to
the formation of bright spatial solitons \cite{Akhmanov72}. The interest in
spatial patterns in lasers was later revived by a series of works. In \cite
{Lugiato88a,Lugiato90}, some nontrivial stationary and dynamic transverse
mode formations in laser beams were demonstrated. It was also recognized 
\cite{Coullet89a} that the laser Maxwell--Bloch equations admit vortex
solutions. The transverse mode formations in \cite{Lugiato88a,Lugiato90},
and the optical vortices in \cite{Coullet89a} were related to one another,
and the relation was confirmed experimentally (Fig. 1.1)\cite
{Tamm88,Brambilla91a}. The optical vortices found in lasers are very similar
to the phase defects in speckle fields reported earlier \cite
{Berry79,Baranova81}. 
\begin{figure}[t]
\includegraphics[width=.7\textwidth]{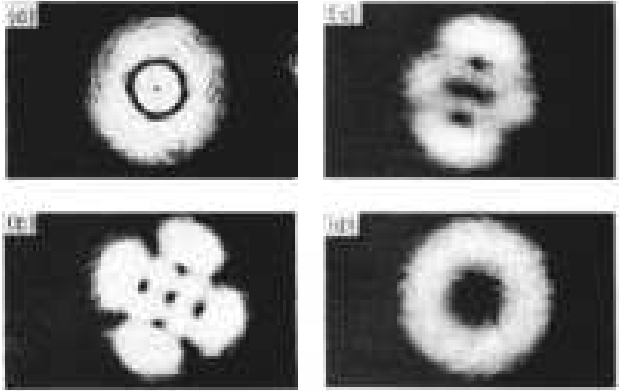}
\caption{The simplest patterns generated by a laser, which can be
interpreted as locked transverse modes of a resonator with curved mirrors.
From \protect\cite{Brambilla91a}, \copyright 1991 American Physical Society}
\label{fig101}
\end{figure}

The above pioneering works were followed by an increasing number of
investigations. Efforts were devoted to deriving an order parameter equation
for lasers and other nonlinear resonators; this would be a simple equation
capturing, in the lowest order of approximation, the main spatio-temporal
properties of the laser radiation. The Ginzburg--Landau equation, as derived
in \cite{Graham70}, is just a very simple model equation for lasers with
spatial degrees of freedom. Next, attempts were made to derive a more
precise order parameter equation for a laser \cite{Coullet89a,Oppo91}, which
led to an equation valid for the red detuning limit. Red detuning means that
the frequency of the atomic resonance is less than the frequency of the
nearest longitudinal mode of the resonator. This equation, however, has a
limited validity, since it is not able to predict spontaneous pattern
formation: the laser patterns usually appear when the cavity is
blue-detuned. Depending on the cavity aperture, higher-order transverse
modes \cite{Brambilla91a} or tilted waves \cite{Jakobsen92} can be excited
in the blue-detuned resonator.

The problem of the derivation of an order parameter equation for lasers was
finally solved in \cite{Staliunas93,Staliunas95a}, where the complex
Swift--Hohenberg (CSH) equation was derived. Compared with the
Ginzburg--Landau equation, the CSH equation contains additional nonlocal
terms responsible for spatial mode selection, thus inducing a pattern
formation instability. Later, the CSH equation for lasers was derived again
using a multiscale expansion \cite{Lega95}. The CSH equation describes the
spatio-temporal evolution of the field amplitude. Also, an order parameter
equation for the laser phase was obtained, in the form of the Kuramoto
--Sivashinsky equation \cite{Wang93}. It is noteworthy that both the
Swift--Hohenberg and the Kuramoto--Sivashinsky equations appear frequently
in the description of hydrodynamic and chemical problems, respectively.

The derivation of an order parameter equation for lasers means a significant
advance, since it allows one not only to understand the pattern formation
mechanisms in this particular system, but also to consider the
broad-aperture laser in the more general context of pattern-forming systems
in nature \cite{Cross93}.

The success in understanding laser patterns initiated a search for
spontaneous pattern formation in other nonlinear resonators. One of the most
extensively studied systems has been photorefractive oscillators, where the
theoretical background was set out \cite{DAlessandro92a}, complicated
structures experimentally observed \cite{Arecchi91a,Hennequin94} and order
parameter equations derived \cite{Staliunas95b}. Intensive studies of
pattern formation in passive, driven, nonlinear Kerr resonators were also
performed \cite{Mandel93,Tlidi93,Lugiato87,Lugiato88b}. Also, the patterns
in optical parametric oscillators received a lot of attention. The basic
patterns were predicted \cite{Oppo94,Longhi96a}, and order parameter
equations were derived in the degenerate \cite{Staliunas95c,deValcarcel96}
and nondegenerate \cite{Longhi96b} regimes. The connection between the
patterns formed in planar- and curved-mirror resonators was treated in \cite
{deValcarcel97}, where an order parameter equation description of weakly
curved (quasi-plane) nonlinear optical resonators was given.

These are just a\ few examples. In the next section, the general
characteristics of nonlinear resonators, and the state of the art are
reviewed.

\section{Patterns in Nonlinear Optical Resonators}

The patterns discussed in the main body of the book are those appearing in
nonlinear optical resonators only. This particular configuration is
characterized by (1) strong feedback and (2) a mode structure, both due to
the cavity. The latter also implies temporal coherence of the radiation.
Thanks to the feedback, the system does not just perform a nonlinear
transformation of the field distribution, where the fields at the output can
be expressed as some nonlinear function of the fields at the input and of
the boundary conditions. Owing to the feedback, the system can be considered
as a nonlinear dynamical system with an ability to evolve, to self-organize,
to break spontaneously the spatial translational symmetry, and in general,
to show its ``own'' distributions not present in the initial or boundary
conditions.

Nonlinear optical resonators can be classified in different ways: by the
resonator geometry (planar or curved), by the damping rates of the fields
(class A, B or C lasers), by the field--matter interaction process (active
and passive systems) and in other ways. After order parameter equations were
derived for various systems, a new type of classification became possible.
One can distinguish several large groups of nonlinear resonators, each of
which can be described by a common order parameter equation:

\begin{enumerate}
\item  Laser-like nonlinear resonators, such as lasers of classes A and C,
photorefractive oscillators, and nondegenerate optical parametric
oscillators. They are described by the complex Swift--Hohenberg equation, 
\begin{equation*}
\frac{\partial A}{\partial t}=\left( D_{0}-1\right) A-A\left| A\right|
^{2}+i\left( a\nabla ^{2}-\omega \right) A-\left( a\nabla ^{2}-\omega
\right) ^{2}A\;,
\end{equation*}
and show optical vortices as the basic localized structures, and tilted
waves and square vortex lattices as the basic extended patterns.

\item  Resonators with squeezed phase, such as degenerate optical parametric
oscillators and degenerate four-wave mixers. They are described, in the most
simplified way, by the real Swift--Hohenberg equation, 
\begin{equation*}
\frac{\partial A}{\partial t}=\left( D_{0}-1\right) A-A^{3}+\left( a\nabla
^{2}-\omega \right) ^{2}A\;,
\end{equation*}
and show phase domains and phase solitons as the basic localized structures,
and stripes and hexagons as the basic extended patterns.

\item  Lasers with a slow population inversion $D$ (class B lasers). They
cannot be described by a single order parameter equation, but can be
described by two coupled equations, 
\begin{eqnarray*}
\frac{\partial A}{\partial t} &=&\left( D-1\right) A+\text{i}\left( a\nabla
^{2}-\omega \right) A-\left( a\nabla ^{2}-\omega \right) ^{2}A\;, \\
\frac{\partial D}{\partial t} &=&-\gamma \left( D-D_{0}+\left| A\right|
^{2}\right) ,
\end{eqnarray*}

and their basic feature is self-sustained dynamics, in particular the
``restless vortex''.

\item  Subcritical nonlinear resonators, such as lasers with intracavity
saturable absorbers or optical parametric oscillators with a detuned pump.
The effects responsible for the subcriticality give rise to additional terms
in the order parameter equation, which in general has the form of a modified
Swift--Hohenberg equation, 
\begin{equation*}
\frac{\partial A}{\partial t}=F\left( D_{0},A,\left| A\right| ^{n},\nabla
^{2}\right) +\text{i}\left( a\nabla ^{2}-\omega \right) A-\left( a\nabla
^{2}-\omega \right) ^{2}A\;,
\end{equation*}
where $F$ represents a nonlinear, nonlocal function of the fields. Its
solutions can show bistability and, as consequence, such systems can support
bistable bright spatial solitons.
\end{enumerate}

This classification is used throughout this book as the starting point for
studies of pattern formation in nonlinear optical resonators. The main
advantage of this choice is that one can investigate dynamical phenomena not
necessarily for a particular nonlinear resonator, but for a given class of
systems characterized by a common order parameter equation, and consequently
by a common manifold of phenomena.

In this sense, the patterns in nonlinear optics can be considered as related
to other patterns observed in nature and technology, such as in
Rayleigh--B\'{e}nard convection \cite{Rayleigh16}, Taylor--Couette flows 
\cite{DiPrima81}, and in chemical \cite{Turing52} and biological \cite
{Meinhardt82} systems. The study of patterns in nonlinear resonators has
been strongly influenced and profited from the general ideas of Haken's
synergetics \cite{Haken77} and Prigogine's dissipative structures \cite
{Prigogine68,Nicolis77}. On the other hand, the knowledge achieved about
patterns in nonlinear resonators provides feedback to the general
understanding of pattern formation and evolution in nature.

Next we review the basic transverse patterns observable in a large variety
of optical resonators. It is convenient to distinguish between two kinds of
patterns: localized structures, and extended patterns in the form of
spatially periodic structures.

\subsection{Localized Structures: Vortices and Solitons}

A transverse structure which enjoys great popularity and on which numerous
studies have been performed, is the optical vortex, a localized structure
with topological character, which is a zero of the field amplitude and a
singularity of the field phase.

Although optical vortices have been mainly studied in systems where free
propagation occurs in a nonlinear material (see Sect. 1.3), some works have
treated the problem of vortex formation in resonators. As mentioned above,
the early studies of these fascinating objects \cite
{Coullet89a,Tamm88,Brambilla91a,Berry79,Baranova81} strongly stimulated
interest in studies of pattern formation in general. The existence of
vortices indicates indirectly the analogy between optics and hydrodynamics 
\cite{Staliunas93,Brambilla91b,Staliunas92,Arecchi91b}. It has been shown
that the presence of vortices may initiate or stimulate the onset of
(defect-mediated) turbulence \cite
{Arecchi91a,Coullet89b,Gil92,Huyet95,Indebetow92}. Vortices may exist as
stationary isolated structures \cite{Neubecker93,Lippi93} or be arranged in
regular vortex lattices \cite{Brambilla91a,Staliunas95a,Hennequin94}. Also,
nonstationary dynamics of vortices have been reported, both of single
vortices \cite{Balzer98,Weiss93} and of vortex lattice structures \cite
{Staliunas95d}. Recently, optical vortex lattices have been experimentally
observed in microchip lasers \cite{Chen02}.

Another type of localized structure is spatial solitons, which are
non-topological structures. Although such structures do not appear
exclusively in optical systems \cite{Gorshkov94,Tsimring97,Dewel95}, they
are now receiving tremendous interest in the field of optics owing to
possible technological applications. A spatial soliton in a dissipative
system, being bistable, can carry a bit of information, and thus such
solitons are very promising for applications in parallel storage and
parallel information processing.

Spatial solitons excited in optical resonators are usually known as cavity
solitons. Cavity solitons can be classified into two main categories:
amplitude (bright and dark) solitons, and phase (dark-ring) solitons.
Investigations of the formation of bright localized structures began with
early work on bistable lasers containing a saturable absorber \cite
{Rosanov90,Rosanov96} and on passive nonlinear resonators \cite{Fauve90}.

Amplitude solitons can be excited in subcritical systems under bistability
conditions, and can be considered as homoclinic connections between the
lower (unexcited) and upper (excited) states. They have been reported for a
great variety of passive nonlinear optical resonators, such as degenerate 
\cite{Staliunas97a,Longhi97,Trillo98} and nondegenerate \cite
{Longhi98b,deValcarcel00} optical parametric oscillators, and for
second-harmonic generation \cite{Etrich97,Longhi98a,Peschel98} (Fig. 1.2),
where the bistability was related to the existence of a nonlinear resonance 
\cite{deValcarcel96}. In some systems, the interaction of solitons and their
dynamical behavior have been studied \cite{Peschel98,Skryabin99a,Skryabin99b}%
. Resonators containing Kerr media also support amplitude solitons, as a
result of either Kerr \cite{Firth96} or polarization (vectorial) \cite
{Sanchez00} instabilities. 
\begin{figure}[b]
\includegraphics[width=.8\textwidth]{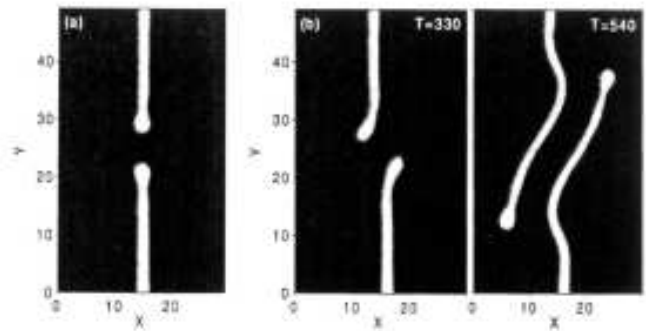}
\caption{Interaction of two moving amplitude solitons in vectorial
intracavity second-harmonic generation: (\textbf{a}) central collision, (%
\textbf{b}) noncentral collision. From \protect\cite{Peschel98}, \copyright
1998 American Physical Society}
\label{fig102}
\end{figure}

In active systems, bright solitons have been demonstrated in photorefractive
oscillators \cite{Saffman94b,Slekys98,Krolikowski98} and in lasers
containing saturable absorbers \cite{Taranenko97,Staliunas98a} or an
intracavity Kerr lens \cite{Dunlop98}. A promising system for practical
applications is the vertical cavity surface emission laser (VCSEL), which
forms a microresonator with a semiconductor as a nonlinear material. The
theoretically predicted patterns for this system \cite
{Brambilla97,Michaelis97,Spinelli98,Tissoni99,Maggipinto00,Spinelli99} were
recently experimentally confirmed in \cite{Taranenko00}.

The required subcriticality condition is usually achieved by introducing an
intracavity absorbing element. However, recently, stable solitons in the
absence of an additional medium have been reported in cascade lasers \cite
{Vilaseca01}.

Besides the amplitude solitons in subcritical nonlinear resonators, a
different type of bistable soliton exists in supercritical resonators. Such
systems are characterized by a broken phase symmetry of the order parameter,
and solutions with only two possible phase values are allowed. In this case
the solitons connect two homogeneous solutions of the same amplitude but of
opposite phase. Such phase solitons, which are round, stable phase domains
of minimum size, appear as a dark ring on a bright background. This novel
type of optical soliton is now receiving a lot of interest, since thes
solitons are seemingly much easier to realize experimentally than their
bright\ counterparts in subcritical systems. 
\begin{figure}[t]
\sidecaption
\includegraphics[width=.4\textwidth]{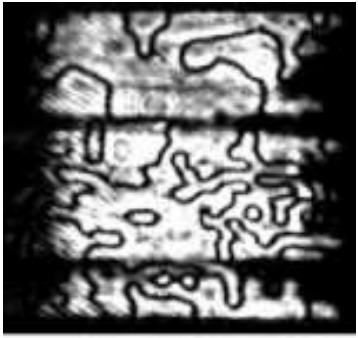}
\caption{Phase domains and phase (dark-ring) solitons in a cavity
four-wave-mixing experiment. From \protect\cite{Weiss99}, \copyright 1999
Optical Society of America}
\end{figure}

One of the systems most investigated has been the degenerate optical
parametric oscillator (DOPO), either in the one-dimensional case \cite
{Longhi96,Trillo97} or in the more realistic case of two transverse
dimensions \cite{Staliunas98b,LeBerre99,Oppo99,Tlidi00a}. Also, the soliton
formation process \cite{LeBerre00,Tlidi00b,Oppo01} and its dynamical
behavior \cite{Izus00,Izus01} have been analyzed. Optical bistability in a
passive cavity driven by a coherent external field is another example of a
system supporting such phase solitons \cite
{Tlidi94,Tlidi97,Tlidi98b,Tlidi98d,Tlidi99b}. Both the DOPO and systems
showing optical bistability are systems described by a common order
parameter equation, the real Swift--Hohenberg equation \cite{Staliunas98c}.
Systems with a higher order of nonlinearity, such as vectorial Kerr
resonators, have also been shown to support phase solitons \cite
{Gallego00,Hernandez00,Gomila01}. 
\begin{figure}[b]
\includegraphics[width=.7\textwidth]{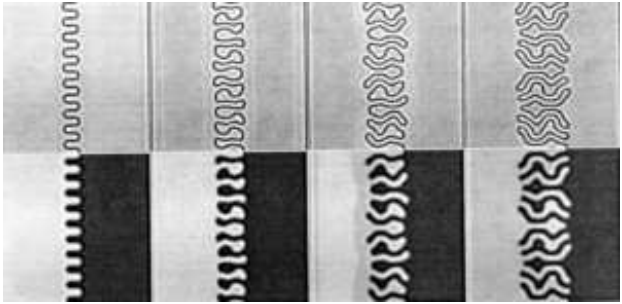}
\caption{Modulational instability of a straight domain boundary and
formation of a finger pattern, in a type II degenerate optical parametric
oscillator. The \textit{upper row} shows the intensity, and the \textit{%
lower row} the phase pattern. From \protect\cite{Izus01}, \copyright 2001
American Physical Society}
\end{figure}

Phase solitons can form bound states, resulting in soliton aggregates or
clusters \cite{Staliunas98b,Vladimirov02}.

Phase solitons in a cavity are seemingly much easier to excite than their
counterparts in subcritical systems. In fact, such phase solitons have
already been experimentally demonstrated in degenerate four-wave mixers \cite
{Taranenko98,Taranenko99,Weiss99} (Figs. 1.3 and 1.4).

\subsection{Extended Patterns}

Besides the localized patterns, vortices and solitons, to which the book is
mainly devoted, extended patterns in optical resonators have been also
extensively studied. In optical resonators, two main categories of patterns
can be distinguished. One class of patterns appears in low-aperture systems,
characterized by a small Fresnel number, such as a laser with curved
mirrors. Since this is the most typical configuration of an optical cavity,
this phenomenon was observed in the very first experimental realizations,
although a systematic study was postponed to a later time \cite{Tamm88}. The
patterns of this kind are induced by the boundary conditions, and can be
interpreted as a weakly nonlinear superposition of a small number of cavity
modes of Gauss--Hermite or Gauss--Laguerre type.

Theoretical predictions based on modal expansions of the field \cite
{Lugiato90,Prati94,Brambilla94} have been confirmed by a large number of
experiments, some of them reported in \cite
{Lin90,Coates94,Li94,Barsella95,Louvergneaux96}. Owing to the particular
geometry of the cavity, this kind of pattern is almost exclusively optical.
If the aperture is increased, the number of cavity modes excited can grow,
and so the spatial complexity of the pattern grows\cite{Chen94}.

The other class of extended optical patterns is typical of large-aperture
resonators, formed by plane mirrors in a ring or a Fabry--P\'{e}rot
configuration. The transverse boundary conditions have a weak influence on
the system dynamics, in contrast to what happens in small-aperture systems.
Consequently, the patterns found in these systems are essentially nonlinear,
and the system dynamics can be reduced to the evolution of a single field,
called the\ order parameter.

The simplest patterns in these systems consist of a single tilted or
traveling wave (TW), which is the basic transverse solution in a laser \cite
{Jakobsen92}, although more complex solutions formed by several TWs have
been found \cite{Huyet94,Feng94}. The predicted laser TW patterns have been
observed in experiments with large-Fresnel-number cavities \cite
{Encinas96,Hegarty99,Huyet01}. The TW solutions are also found in passive
resonators described by the same order parameter equation, such as
nondegenerate optical parametric oscillators (OPOs) \cite
{Longhi96a,Sanchez98}. The effect of an externally injected signal in a
laser has been also studied \cite{Georgiou94,Longhi97b}, showing the
formation of more complex patterns, such as rolls or hexagons.

Roll, or stripe, patterns are commonplace for a large variety of nonlinear
passive cavities, such as degenerate OPOs \cite{Oppo94}, four-wave mixers 
\cite{deValcarcel96}, systems showing optical bistability \cite
{Tlidi93,Nalik92} and cavities containing Kerr media \cite{Geddes94a}.
Patterns with hexagonal symmetry are also frequently found in such
resonators \cite{Firth92,Taranenko02}. Both types of pattern are familiar in
hydrodynamic systems, such as systems showing Rayleigh--B\'{e}nard
convection.

Another kind of traveling solution existing in optical resonators
corresponds to spiral patterns, such as those found in lasers \cite
{Yu96,Yu99} and in OPOs \cite{Lodahl00,Longhi01a}, which are typical
structures in chemical reaction--diffusion systems.

When more complex models, including additional effects are considered, a
larger variety of patterns, sometimes of exotic appearance, is found. Some
such models generalize the above cited models by considering the existence
of competition between different parametric processes \cite
{Lodahl99,Longhi01b} or between scalar and vectorial instabilities \cite
{Hoyuelos02}, the walk-off effect due to birefringence in the medium \cite
{Santagiustina97,Ward98,Santagiustina98}, or external temporal variation of
the cavity parameters \cite{Longhi00}.

Some systems allow the simultaneous excitation of patterns with different
wavenumbers. These systems form patterns with different periodicities that
have been called quasicrystals \cite{Longhi99a,Longhi99b} and daisy patterns 
\cite{LeBerre96} (Fig. 1.5). 
\begin{figure}[b]
\sidecaption
\includegraphics[width=.5\textwidth]{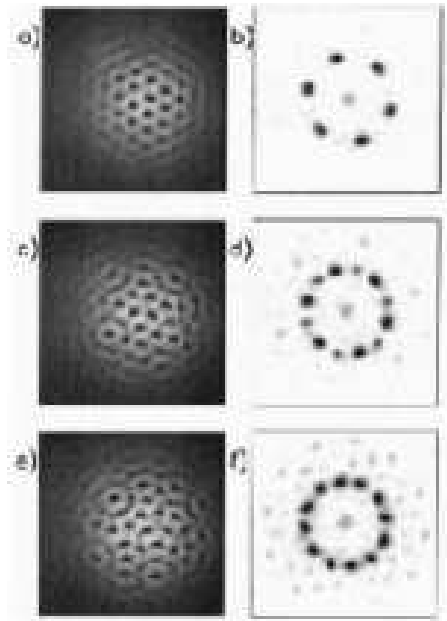}
\caption{Experimentally observed hexagonal patterns with sixfold and
twelvefold symmetry (quasipatterns), in a nonlinear optical system with
continuous rotational symmetry. From \protect\cite{Herrero99}, \copyright
1999 American Physical Society}
\end{figure}

The experimental conditions for large-aspect-ratio resonators are not easy
to achieve. Most of the experiments performed have studied multimode regimes
involving high-order transverse modes. The formation of the patterns
described above was reported in lasers \cite{Encinas96,Hegarty99,Huyet01}
and OPOs \cite{Vaupel99,Ducci01}. The observed patterns correspond well to
the numerical solutions of large-aspect-ratio models. Conditions for
boundary-free, essentially nonlinear patterns were obtained in \cite
{Saffman94b,Staliunas97b} with the use of self-imaging resonators, which
allowed the experimenters to obtain Fresnel numbers of arbitrarily high
value.

All the patterns reviewed above are two-dimensional, the light being
distributed in the transverse space perpendicular to the resonator axis, and
evolving in time. Recently, the possibility of three-dimensional patterns
was demonstrated for OPOs \cite{Staliunas98d}, nonlinear resonators with
Kerr media \cite{Tlidi98a,Tlidi00c}, optical bistability \cite{Tlidi98c} and
second-harmonic generation \cite{Tlidi99}.

Finally, the problem of the effect of noise on the pattern formation
properties of a nonlinear resonator has also been treated. One can expect
that noise, which is present in every system, will bring about new features
in the spatio-temporal dynamics of the system. First, noise can modify
(shift) the threshold of pattern formation \cite{GarciaOjalvo99}. Second,
owing to noise, the precursors of patterns can be seen below the pattern
formation threshold \cite{Deissler89,GarciaOjalvo93,Mueller97}. While a
noiseless pattern-forming system below the pattern formation threshold shows
no pattern at all, since all perturbations decay, one observes in the
presence of noise a particular form of spatially filtered noise, which in
the field of nonlinear optics has been called ``quantum patterns'' when the
noise is of quantum origin \cite{Lugiato93,Gatti97,Zambrini00}. Above the
pattern formation threshold, noise can also result in defects (dislocations
or disclinations) of the patterns \cite{Vinals91,Elder92}.

\section{Optical Patterns in Other Configurations}

In parallel with the studies on nonlinear resonators, pattern formation
problems have been considered in other optical configurations. These
configurations can be divided into the following categories, according to
teir geometry and complexity.

\subsection{Mirrorless Configuration}

When an intense light beam propagates in a nonlinear medium, it can
experience filamentation effects, leading to periodic spatial distributions 
\cite{Mamaev96b}, or develop into self-trapped states of light, or solitons.
The self-focusing action of the nonlinearity compensated by diffraction
results in self-sustained bright spatial solitons \cite{Akhmanov72}, which
can exist as isolated states or form complex ensembles, sometimes
interacting in a particle-like fashion \cite
{Hayata93,Torruellas95,Mihalache00,Ripoll00,Skryabin98,Mitchell98,Stegemann99}%
. Also, dark solitons \cite
{Kivshar94,Haelterman94a,Haelterman94b,Sheppard94,Haelterman96,Mamaev96a,Neshev97,Dreischuh99,Dreischuh02}
and optical vortices \cite
{Basistiy93,Swartzlander92,Swartzlander93,Swartzlander94,Gahagan96,Swartzlander97a,Swartzlander97b,Swartzlander98,Swartzlander99,Rozas00,Kivshar01,Luther97}
have been described and experimentally observed. In such a mirrorless
configuration feedback is absent, and one obtains not a spontaneous pattern
formation, but just a nonlinear transformation of the input distribution.
This nonlinear transformation can be very complicated, and can be described
by complicated integro-differential equations. However, every transformation
remains a transformation, and without feedback it does not lead to
spontaneous pattern formation. Some other mirrorless schemes, where optical
pattern formation has been predicted, are based on the interaction of two
counterpropagating pumping waves in a nonlinear medium. It has been shown
that the waves that appear through nonlinear mixing processes have their
lowest threshold at certain angles with respect to the pumping waves, and
may result in a wide variety of patterns, either extended, such as rolls or
hexagons \cite
{Firth88,Grynberg88,Grynberg89,Haelterman93,Saffman94,Geddes94b,Lushnikov99,Schwab01}%
(Fig. 1.6), or localized \cite{Pitois98}. Experimental confirmation has been
obtained using various nonlinear media, such as atomic vapors and
photorefractive crystals.\vspace{0.2cm} 
\begin{figure}[h]
\sidecaption
\includegraphics[width=.6\textwidth]{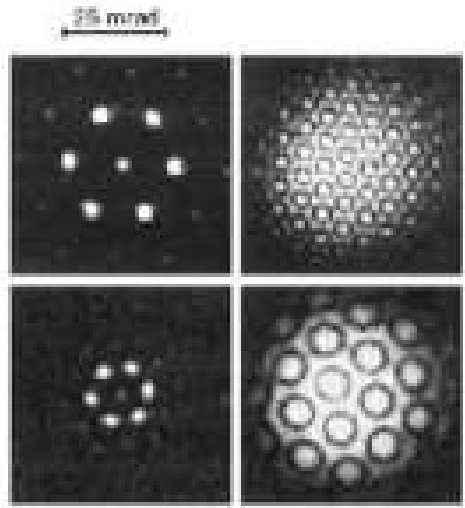}
\caption{Hexagonal patterns with different spatial scales observed in a
photorefractive crystal with a single pump wave. From \protect\cite
{Lushnikov99}, \copyright 1991 Optical Society of America}
\end{figure}

\subsection{Single-Feedback-Mirror Configuration}

The presence of a mirror introduces feedback into the system. Unlike the
case in the previous schemes, here nonlinearity and diffraction act at
different spatial locations. The most typical configuration is formed by a
thin slice of a Kerr medium and a mirror at some distance. Theoretical
studies have predicted structures mainly with hexagonal symmetry \cite
{Firth91,DAlessandro92b,DAlessandro95,Scroggie96} (Fig. 1.7), although more
complex solutions have been found \cite{Vorontsov97a,Denz98}. From the
experimental side, various nonlinear media, such as atomic vapors \cite
{Giusfredi88,Lange96}, and Kerr \cite{Vorontsov97b} and photorefractive \cite
{Denz98} media have been used successfully. Also, this configuration led to
the first realization of localized structures in nonlinear optics \cite
{Kreuzer96}. The dynamics and interaction of these localized structures have
been extensively investigated \cite
{Schreiber97,Kreuzer98,Logvin00,Schapers00} (Fig. 1.8). 
\begin{figure}[h]
\includegraphics[width=.8\textwidth]{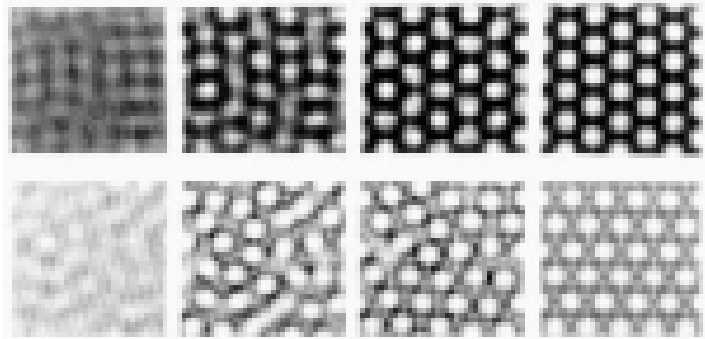}
\caption{Hexagon formation in a single-feedback-mirror configuration.
Numerical results from \protect\cite{DAlessandro92b}, \copyright 1991
American Physical Society}
\end{figure}
\begin{figure}[h]
\includegraphics[width=.99\textwidth]{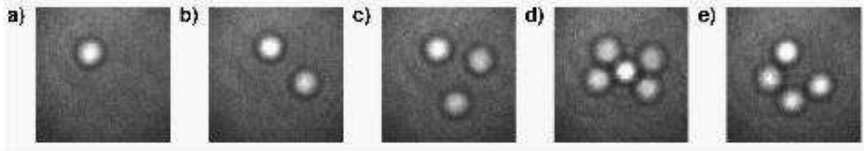}
\caption{Dissipative solitons observed experimentally in sodium vapor with a
single feedback mirror. From \protect\cite{Schapers00}, \copyright 2000
American Physical Society}
\end{figure}

\subsection{Optical Feedback Loops}

Another configuration, somewhat between the single feedback mirror and the
nonlinear resonator, is the feedback loop. In such a configuration, one has
the possibility of acting on the field distribution on every round trip
through the loop, continuously transforming the pattern distribution. Some
typical two-dimensional transformations are the rotation, translation,
scaling and filtering of the pattern. The first work obtained pattern
formation by controlling the spatial scale and the topology of the
transverse interaction of the light field in a medium with cubic
nonlinearity \cite{Akhmanov92,Vorontsov94a,Zheleznik94}, by controlling the
phase of the field with a spatial Fourier filter \cite
{Vorontsov94b,Vorontsov98} (Figs. 1.9 and 1.10), and by introducing a medium
with a binary-type refractive nonlinear response \cite{Samson97}. 
\begin{figure}[t]
\sidecaption
\includegraphics[width=.5\textwidth]{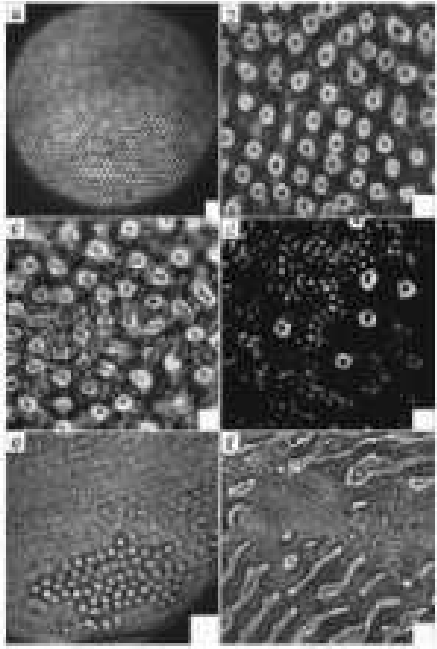}
\caption{Experimental patterns in an optical system with two-dimensional
feedback. (\textbf{a}) Hexagonal array, (\textbf{b})--(\textbf{d})
``black-eye'' patterns, (\textbf{e}) island of bright localized structures, (%
\textbf{f}) optical squirms. From \protect\cite{Vorontsov98}, \copyright
1998 American Physical Society}
\end{figure}
\begin{figure}[t]
\sidecaption
\includegraphics[width=.5\textwidth]{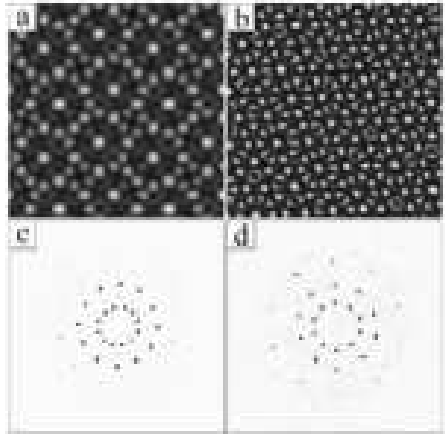}
\caption{Quasicrystal patterns with dodecagonal symmetry, with different
spatial scales, together with the corresponding spatial spectra. From 
\protect\cite{Vorontsov98}, \copyright 1998 American Physical Society}
\end{figure}

A very versatile system is a feedback loop with a liquid-crystal light valve
acting as a phase modulator with a Kerr-type nonlinearity. The conversion
from a phase to an intensity distributiocn, required to close the feedback
loop, can be performed by two means: by free propagation (diffractional
feedback) \cite{Pampaloni93,Pampaloni94} or by interference with reflected
waves (interferential feedback), as shown in Fig. 1.11 \cite
{Arecchi95,Ramazza96b,Ramazza98}. In both cases, a great variety of
kaleidoscope-like patterns have been obtained theoretically and
experimentally. The patterns can also be controlled by means of nonlocal
interactions, via rotation \cite{Pampaloni95,Residori96,Pampaloni97} (Fig.
1.12) or translation \cite{Ramazza96a,Ramazza97} of the signal in the
feedback loop, giving rise to more exotic solutions such as quasicrystals
and drifting patterns. The existence of spatial solitons and the formation
of bound states of solitons have also been reported experimentally in the
liquid-crystal light valve system \cite{Ramazza02}, as shown in Fig. 1.13. 
\begin{figure}[t]
\includegraphics[width=.99\textwidth]{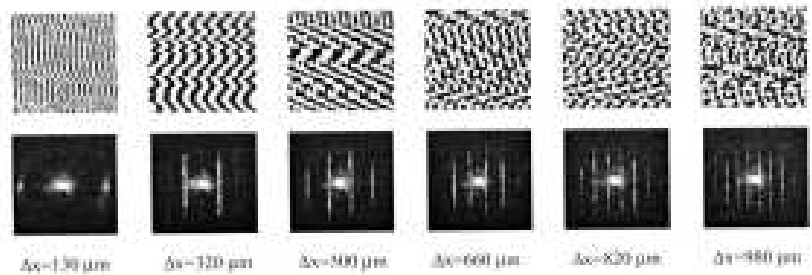}
\caption{Patterns in a liquid-crystal light valve in the interferential
feedback configuration, for increasing translational nonlocality $\varDelta %
x $. The near field (\textit{top row}) is shown together with the
corresponding spectrum (\textit{bottom row}). From \protect\cite{Ramazza98},
\copyright 1998 American Physical Society}
\end{figure}
\begin{figure}[t]
\includegraphics[width=.9\textwidth]{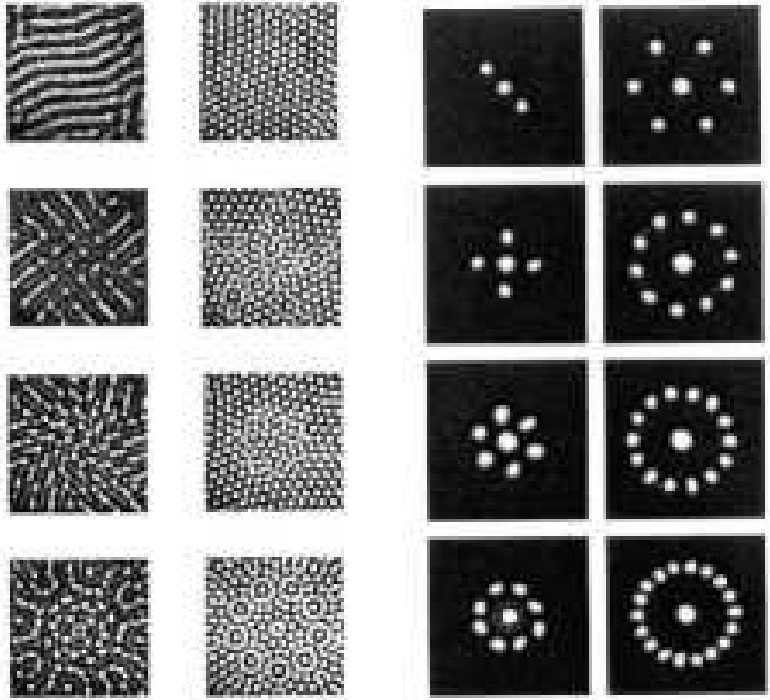}
\caption{Crystal and quasicrystal patterns obtained experimentally by
rotation of the signal in a liquid-crystal light valve feedback loop. The 
\textit{first} and \textit{second columns} show the near-field
distributions, and the \textit{third} and \textit{fourth columns} the
corresponding far fields. From \protect\cite{Pampaloni95}, \copyright 1995
American Physical Society}
\end{figure}
\begin{figure}[h]
\sidecaption
\includegraphics[width=.3\textwidth]{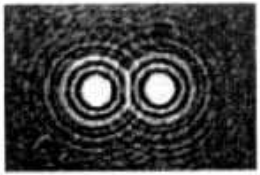}
\caption{Bound state of spatial solitons in a liquid-crystal light valve
interferometer. From \protect\cite{Ramazza02}, \copyright 2002 American
Physical Society}
\end{figure}

\section{The Contents of this Book}

In Chaps. 2 and 3, the order parameter equations for broad-aperture lasers
and for other nonlinear resonators are obtained. These chapters are
relatively mathematical; however, the OPEs derived here pave the way for the
subsequent chapters of the book. The derivation of the OPEs for class A and
class C lasers is given in Chap. 2. For completeness, two techniques of
derivation are given: one based on the adiabatic elimination of the fast
variables, and one based on multiscale expansion techniques. Both procedures
lead to the complex Swift--Hohenberg equation as the OPE for lasers. The CSH
equation describes the spatio-temporal dynamics of the complex-valued order
parameter, which is proportional to the envelope of the optical field. In
Chap. 3, the OPEs for optical parametric oscillators and photorefractive
oscillators (PROs) are derived. In the degenerate case, the resulting
equation is shown to be the real Swift--Hohenberg equation, first obtained
in a hydrodynamic context. For large pump detuning values, a generalized
model including nonlinear resonance effects is obtained. In the case of
PROs, the adiabatic elimination technique is used to derive the CSH
equation. The order parameter equations derived in Chaps. 2 and 3 divide
nonlinear optical resonators into distinct classes, and thus allow one to
study pattern formation phenomena without necessarily considering every
nonlinear optical system separately; instead, one can consder classes of the
systems.

Chapters 4 and 5 are devoted to the patterns of the first class of systems,
that described by the CSH equation, i.e. lasers, photorefractive oscillators
and nondegenerate OPOs. The localized patterns in this class of systems are
optical vortices: these are zeros of the amplitude of the optical field, and
are simultaneously singularities of the field phase. Optical vortices
dominate the dynamics of the system in near-resonant cases (when the
detuning is close to zero). The CGL equation in this near-resonant limit can
be rewritten in a hydrodynamic form. Owing to this analogy between laser and
hydrodynamics, the dynamics of the transverse distribution of the laser
radiation are very similar to the dynamics of a superfluid. It is shown that
optical vortices of the same topological charge rotate around one another; a
pair of vortices of the same charge translate in parallel through the
aperture of the laser or annihilate, depending on the parameters.

In Chap. 5, the limit of large or moderate detuning is considered. The CSH
equation cannot be rewritten in a hydrodynamic form, but the dynamics of
the\ fields can still be well interpreted by hydrodynamic means. For large
detuning, tilted waves are excited. In hydrodynamic terms, flows with a
velocity of fixed magnitude but arbitrary direction are favored. This
results, in particular, in counterpropagating flows separated by vortex
sheets. This also leads to optical vortices advected by the mean flow, and
similar phenomena. Such phenomena are analyzed theoretically and
demonstrated numerically. A pattern of square symmetry, called a square
vortex lattice, consisting of four counterpropagating flows in the form of a
cross, is also described and discussed.

In Chap. 6, the effects of the curvature of the mirrors of the resonator are
analyzed. The majority of theoretical investigations of pattern formation in
nonlinear optics, including those in the largest part of this book, have
been performed by assuming a plane-mirror cavity model. However, in
experiments resonators with curved mirrors are often used. Therefore a model
of a laser with curved mirrors is introduced. The presence of curved mirrors
results in an additional term in the order parameter equation, proportional
to the total curvature of the mirrors in the resonator. This term produces a
coordinate-dependent (parabolic) phase shift of the order parameter during
propagation in the resonator. The presence of the parabolic potential allows
one to expand the field of the resonator in terms of the eigenfunctions
(transverse modes) of the potential. Although this mode expansion is
strictly valid for linear resonators only, the nonlinearity in the resonator
results in a weakly nonlinear coupling of the complex amplitudes of the
modes. As a result, an infinite set of coupled ordinary differential
equations for complex-valued mode amplitudes is derived. This gives an
alternative way of investigating the transverse dynamics of a laser, by
solving the equations for the mode amplitudes instead of solving the partial
differential equations. The technique of mode expansion is shown to be
extremely useful when one is dealing with a small number of transverse
modes. In particular, the transverse dynamics of class A lasers and
photorefractive oscillators are considered; the phenomena of transverse mode
pulling and locking are observed. Chapter 6 also deals with degenerate
resonators, such as self-imaging and confocal resonators. In such
resonators, the longitudinal mode separation is an integer multiple of the
transverse mode separation. It is shown, by analysis of the corresponding 
\textit{ABCD} matrices, that self-imaging resonators are equivalent to
planar resonators of zero length. This insight opened up new possibilities
for experimenting with transverse patterns in nonlinear optical systems, and
allowed the first experimental realization of a number of phenomena
predicted theoretically for nonlinear resonators.

Chapter 7 deals with patterns in class B lasers. Class B lasers are not
describable by the CSH equation. Owing to the slowness of the population
inversion, the order parameter equation in this case is not a single
equation belonging to one of the classes defined above, but a system of two
coupled equations, resembling those derived for excitatory or oscillatory
chemical systems, where the (slow) population inversion plays the role of
the recovery variable, and the fast optical field plays the role of the
excitable variable. An analysis of such self-sustained spatio-temporal
dynamics in a class B laser is performed. The vortices, which are stationary
in a class A laser, perform self-sustained meandering in a class B laser, a
phenomenon known as the ``restless vortex''. Also, vortex lattices
experience self-sustained oscillatory dynamics. Either the vortices in the
lattice oscillate in such a way that neighboring vortices rotate in
antiphase, thus resulting in an ``optical'' mode of vortex lattice
oscillation, or the vortex lattice drifts spontaneously with a well-defined
velocity, thus resulting in an ``acoustic'' oscillation mode.

The following chapters, Chaps. 8 to 11, are devoted to amplitude and phase
domains, as well as amplitude and phase solitons in bistable nonlinear
optical systems. The general theory of subcritical spatially extended
systems is developed in Chap. 8, where two mechanisms of creation of
subcriticality in optical resonators are described: one due to the presence
of a saturable absorber, and one due to the presence of a nonlinear
resonance. A discussion in terms of order parameter equations is given.

In Chap. 9, a theoretical description and experimental evidence of domain
dynamics and spatial solitons in lasers containing a saturable absorber are
presented. Two different resonator configurations are used: a self-imaging
resonator where both nonlinearities (due to the gain and to saturable
absorption) are placed at the same location on the optical axis of the
resonator, and a self-imaging resonator where the two nonlinearities are
placed at Fourier-conjugated locations. For spatially coincident
nonlinearities, the evolution of domains is demonstrated numerically and
experimentally, with the eventual appearance of spatial solitons. For
nonlinearities placed in conjugate locations in the resonator, the
competition, mutual interaction and drift of solitons are investigated, also
both theoretically and experimentally.

In Chap. 10, a subcriticality mechanism different from saturable absorption
is studied, in this case related to the existence of a nonlinear resonance
due to nonresonant pumping. As an example, the order parameter equation
obtained in Chap. 3 for a degenerate OPO with a detuned pump is considered.
The nonlinear resonance implies that the pattern wavenumber depends on the
intensity of the radiation. With approriate values of the detuning, the
nonlinear resonance can lead to bistability, and thus allow the excitation
of amplitude domains and spatial solitons. Numerical results from the DOPO
mean-field model are given for comparison.

In Chap. 11, the dynamics of phase domains in supercritical real-valued
order parameter systems, such as the degenerate OPO, are analized. These
systems should properly be described by the real Swift--Hohenberg equation.
It is demonstrated that the domain boundaries, the lines of zero intensity
separating domains of opposite phase, may contract or expand depending on
the value of the resonator detuning. In this way, the domain boundaries
behave as elastic ribbons, with the elasticity coefficient depending on the
detuning. Contracting domains, observed for small values of the detuning,
eventually disappear. Expanding domains are found for large values of the
detuning, and their evolution results in labyrinthine structures. For
intermediate values of the detuning, the contracting domain boundaries stop
contracting at a particular radius. The latter scenario results in stable
rings of domain boundaries, which are phase solitons. The experimental
confirmation of the predicted phenomena is described.

In Chap. 12, the Turing pattern formation mechanism, typical of chemical
reaction--diffusion systems, is shown to exist also in nonlinear optics. The
pattern formation mechanism described in most of the chapters of the book is
based on an off-resonance excitation. The Turing mechanism, however, is
based on the interplay between the diffusion and/or diffraction of\
interacting components. In particular, the emergence of Turing-like patterns
is predicted to occur in active and passive systems, concrete examples being
lasers with a strongly diffusing population inversion, and degenerate OPOs
with a strongly diffracting pump wave. In both cases, one field plays the
role of activator, and the other the role of inhibitor. It is also shown
that the effect of diffusion and/or diffraction contributes to the
stabilization of spatial solitons and allows the existence of complex states
resembling molecules of light.

In Chap. 13, we describe the three-dimensional structures of light predicted
to occur in resonators described by the three-dimensional Swift--Hohenberg
equation. This order parameter equation describes a class of nonlinear
optical resonators including the synchronously pumped OPO. Various
structures embedded in the envelopes of spatio-temporal light pulses are
discussed, in the form of extended patterns (lamellae and tetrahedral
patterns), light bubbles (the analogue of the phase solitons in two
dimensions) and vortex rings. These structures exist when the OPO resonator
length is matched to the length of the pump (mode-locked) laser, which emits
a continuous or finite train of picosecond pulses. A three-dimensional
modulation can develop on the subharmonic pulses generated, depending on
several parameters such as the detuning from the resonance of the OPO
cavity, and the mismatch of the resonator lengths for the pump and OPO
lasers.

The final chapter, Chap. 14, deals with the influence of noise on spatial
structures in nonlinear optics. Noise, which is not considered in the rest
of the book, is always present in a real experiment, in the form of vacuum
noise (always inevitable) or noise due to technological limitations. It is
shown that the noise affects the pattern formation in several ways. Above
the modulation instability threshold, where extended patterns are expected,
the noise destroys the long-range order in the pattern. Rolls and other
extended structures still exist in the presence of noise, but they may
display defects (such as dislocations and disclinations) with a density
proportional to the intensity of the noise. Also, below the modulation
instability threshold, where no patterns are expected in the ideal
(noiseless) case, the noise is amplified and can result in (noisy) patterns.
The symmetry of a pattern may show itself even below the pattern formation
threshold, thanks to the presence of noise. This can be compared with a
single-transverse-mode laser, where the coherence in the radiation develops
continuously, and where the spectrum of the luminescence narrows
continuously when the generation threshold is approached\ from below.

\end{document}